\documentclass[preprint,showpacs,preprintnumbers,amsmath,amssymb,aps,floatfix]{revtex4}

\usepackage{dcolumn}% Align table columns on decimal point
\usepackage{graphicx}% Include figure files
\usepackage{epsf}
\def\LL{\left\langle}	% left angle bracket
\def\RR{\right\rangle}	% right angle bracket
\def\LP{\left(}		% left parenthesis
\def\RP{\right)}	% right parenthesis
\def\PAR#1#2{ {\frac{\partial #1}{\partial #2}} }

\newcommand{\BE}{\begin{displaymath}}
\newcommand{\EE}{\end{displaymath}}
\newcommand{\BNE}{\begin{equation}}
\newcommand{\ENE}{\end{equation}}
\newcommand{\BEA}{\begin{eqnarray}}
\newcommand{\EEA}{\nonumber\end{eqnarray}}

\begin{document}

\title{The strange quark condensate in the nucleon in 2+1 flavor QCD}

\author{D. Toussaint and W. Freeman [MILC Collaboration]}
\affiliation{Department of Physics, University of Arizona, Tucson, AZ 85721, USA}

\date{\today}

\begin{abstract}
We calculate the ``strange quark content of the nucleon'',
$\LL N | \bar s s | N\RR$, which is important for interpreting
the results of some dark matter detection experiments.  The method
is to evaluate quark-line disconnected correlations on the MILC
lattice ensembles, which include the effects of dynamical light and strange
quarks.  After continuum and chiral extrapolations, the result is
%$\LL N | \bar s s | N\RR = 0.69(0.07)_{\rm statistical}(0.09)_{\rm systematic}$,
$\LL N | \bar s s | N\RR = 0.69(7)_{\rm statistical}(9)_{\rm systematic}$,
in the modified minimal subtraction scheme (2 GeV), or for the renormalization scheme invariant
form, $m_s \PAR{M_N}{m_s} = 59(6)(8)$ MeV.
\end{abstract}

\maketitle

%\section{Introduction and motivation}

In addition to its relevance to nuclear structure, the strange quark
condensate in the nucleon,
$\LL N | \bar s s|  N \RR$, is important in understanding experimental
searches for dark matter, since some of the leading candidates for
dark matter couple most strongly to the nucleon through
interactions with the strange quark loops.  For example,  the importance
of this quantity is emphasized in Refs.~\cite{BALTZ06,ELLIS08}.
%
% other approaches and other lattice studies
The strange and light quark condensates in the nucleon have been calculated
through effective field theories of nucleons and mesons\cite{NUCEFF},
and the heavy quark content can be studied perturbatively\cite{HEAVY}.
Previous lattice studies  of the nucleon strange quark content have
been done in the quenched approximation\cite{FUKUGITA02,DONG02}, with two flavors of
Wilson or overlap quarks\cite{SESAM98,UKQCD01,JLQCD08} or an exploratory study
with 2+1 flavor stout quarks\cite{BALI08}.
A recent study uses 2+1 flavor baryon mass fits\cite{YOUNGTHOMAS09}.
Some, though by no means all, of these studies have suggested that
$\LL N | \bar s s | N \RR$ might be much larger than found here.
%DT would like to say "natural size", but then have to explain it
% ala Shifman ...

% MILC ensembles, 2+1 sea quarks, larger lattices, check continuum and chiral extrp.
In this work we use lattices with 2+1 flavors of dynamical quarks (two light, one strange)
and nucleon correlators generated from them. These were generated by the MILC
collaboration, except for one long ensemble from the UKQCD collaboration.
These simulations use a Symanzik improved gauge action
%\cite{SYMANZIK}
and an improved staggered quark action.
%\cite{ASQTAD}.
Details of the action, the ensembles of gauge configurations,
and the techniques for
computing the nucleon correlators are in Ref.~\cite{RMP}.
The simulations cover a range of light quark masses and a range
of lattice spacings, which allows us to check the extrapolations
to zero lattice spacing and to the physical light quark mass.
The lattices have a spatial size of 2.4 fm or larger, with
$m_\pi L$ ranging from 3.8 to 6, so that finite size effects
will be small.  Each ensemble, or set
of gauge configurations with a given gauge coupling and quark masses,
used here contains from 500 to 4500 equilibrated lattices, with a total
of 25784 lattices used in the analysis.
%Table~\ref{RUNTABLE} shows the parameters of the lattices used in
%this study.
%\\LEAVE THIS OUT IF WE ARE SHORT ON ROOM\\

%\section{Methods and lattice data}

% relate to deriv of M_N
% relate to deriv of propagator
%REVISE The Feynman-Hellman theorem relates the matrix element
Differentiation of a path integral expression for the nucleon
mass with respect to the strange quark mass (the Feynman-Hellman theorem) relates the matrix element
$\LL N \left| \bar s s \right| N\RR$ to $\PAR{M_N}{m_s}$.
In particular,
\BNE
\LL N \left| \int\, d^3x\, \bar s s \right| N\RR - \LL 0 \left| \int\, d^3x\, \bar s s \right| 0\RR
= \PAR{M_N}{m_s}\Big|_{\alpha_s,m_l}
\ \ \ \ ,\ENE
where the left hand side makes definite what we mean by $\LL N|\bar s s | N \RR$.
Note the vacuum subtraction  and the integral over space.
%We emphasize that the derivative is to be taken with all parameters
%in the lattice action other than $m_s$ held fixed.
%
%% leave out some steps in a letter
%To be specific, in a box with volume $V$ and Euclidean time extent $T$,
%\BNE \label{eq_matdef} \LL N \bar s s N\RR = \frac{e^{M_N T}}{T} \int\,d^3x\,dt\,  \LB \LL N(0) \bar s s (\vec x,t) N(T) \RR
%-  \LL N(0) N(T) \RR \LL \bar s s \RR
%\RB \ \ \ \ ,\ENE
%where we have averaged over time.
%Here $N$ is a normalized zero-spatial-momentum nucleon creation operator, so that
%\BNE \label{eq_nucmass1} \LL N(0) N(T) \RR = e^{-M_N T} \ \ \ \ .\ENE
%The $e^{M_N T}$ in Eq.~\ref{eq_matdef} cancels the
%$e^{-M_N T}$ from propagation of the nucleon from $0$ to $t$ and then from $t$ to $T$.
%We have subtracted the disconnected (vacuum) $\bar s s$.
%The action is $S_{other} + \int \,d^4x \,m_s \bar s s$, where $S_{other}$ is
%all the
%parts not involving the strange quark.   Thus a derivative of $e^{-S}$ with respect to $m_s$
%brings out a factor of $\int \,d^4x \,\bar s s$.
%From differentiating Eq.~\ref{eq_nucmass1} with respect to $m_s$ we find
%\BNE \LL N \int\,d^3x\, \bar s s N \RR = \PAR{M_N}{m_s} \ENE
% end "leave out"

Since we expect that the nucleon is made mostly from light quarks,
%it may seem strange at first to suppose that the mass of the nucleon depends strongly on the
it may seem strange to suppose that $M_N$ depends strongly on the
mass of the strange quark, $m_s$.
%REVISE
%However, the prescription given by the Feynman-Hellman
%theorem is to measure $\PAR{M_N}{m_s}$ with all other parameters in
However, to equate
$\LL N | \bar s s | N \RR$ with $\PAR{M_N}{m_s}$, differentiation of the path
integral must be done with all other parameters in
the action held fixed.
This change in $m_s$ would cause all dimensionful
QCD quantities to change by roughly the same factor,
with most of this change interpreted as a change
in the physical lattice spacing.   For example, if $f_\pi$ were used to determine the
lattice spacing, both lattice quantities $aM_N$ and $af_\pi$ might change, with
ratio $aM_N/af_\pi$ approximately constant.
%Thus it is neither terribly
%surprising nor alarming to find $\PAR{M_N}{m_s}$ of order one.
Thus it is not terribly
surprising to find $\PAR{M_N}{m_s}$ of order one.

% relate to correlation with PBP
%A straightforward way to find $\PAR{M_N}{m_s}$ would be to run simulations at
%different strange quark masses, and subtract to find this derivative,
%or, more generally, fitting the nucleon mass for sea and
%quark masses and gauge couplings to a function
%of parameters including the strange sea quark mass.
%%and differentiating the fitting function.
%WF A straightforward way to find $\PAR{M_N}{m_s}$ would be to run several simulations
%WF changing only the strange quark mass,
%WF and use the change in $M_N$ to compute the derivative.
%WF More generally, one can fit values for $M_N$ to a function of
%WF all the lattice parameters and examine the dependence of the fit form on $m_s$.
%WF %
%WF Since the MILC ensembles contain runs with different
%WF strange quark masses, this method is in principle possible for us.
%WF The fact that these ensembles generally have different gauge couplings for
%WF different quark masses complicates the fitting.
%WF We find that looking at correlations of $\bar s s$ with the
%WF nucleon correlator gives a better signal, although, in one case to be discussed
%WF later, we have a useful check from nucleon mass fits.
The MILC ensembles contain runs with different values for $m_s$,
so it is in principle possible to determine $\PAR{M_N}{m_s}$ from a fit to
$M_N$ on different ensembles. However, we find that correlations of $\bar s s$ with the
nucleon correlator give a better signal.
%In one case to be discussed later, we do  have a useful check from nucleon mass fits.
(In one case to be discussed later, we do  have a useful check from mass fits.)

In the lattice simulations, the nucleon mass $M_N$ is obtained by a fit to the nucleon
%REVISE
%correlator $P(t)$ and as such is just a complicated
%function of the correlator at different times:
correlator $P(t) = \LL {\cal O}_N (0) {\cal O^\prime}_N (t) \RR$,
where ${\cal O}_N$ and ${\cal O^\prime}_N$ are lattice operators with
the best practicable overlap with the nucleon.
As such, it is just a complicated
function of the correlator at different times:
\BNE \label{eq_massfunction1} M_N = f\LP P(t_1),P(t_2),P(t_3) \ldots \RP \ \ \ \ .\ENE
Using the chain rule to rewrite the derivative:
\BNE \PAR{M_N}{m_s} = \sum_i \PAR{M_N}{P(t_i)} \PAR{P(t_i)}{m_s} \ \ \ \ .\ENE

The partial derivatives ${\partial P(t_i)}\over{\partial m_s}$ can be evaluated
by using the Feynman-Hellman
theorem in reverse to relate them to
$\LL P(t_i) \int d^4x\, \bar s s \RR - \LL P(t_i) \RR \LL \int d^4x\, \bar s s \RR$.
%Then the partial derivatives ${\partial M_N}\over{\partial P(t_i)}$ can be evaluated
Then $\PAR{M_N}{m_s}$ is evaluated
by adding and subtracting a small multiple of $\PAR{P(t)}{m_s}$
to the nucleon correlator $P(t)$ and examining the change in the fit result.
It may seem that this second use of the Feynman-Hellman theorem
has just reversed the original calculation relating $\LL N | \bar s s | N \RR$ to
$\PAR{M_N}{m_s}$.  If the source and sink for the lattice nucleon correlator
%$P(t)$ were an operator which created nothing but a normalized nucleon
$P(t)$ created nothing but a normalized nucleon
state, this would be the case.  But in practice the lattice correlator contains
opposite parity particles with almost the same amplitude as the nucleon but with higher mass, and
excited states of both parities.  The fitting procedure implicit in Eq.~\ref{eq_massfunction1} is designed
to determine a nucleon mass from this complicated correlator, by explicitly including opposite parity
(alternating in $t$) contributions and by ignoring the correlator at short separations, so
that the excited state contributions are suppressed.

Nucleon correlators have been computed on most of the MILC
ensembles.   Typically these are averaged over eight Coulomb gauge wall
sources in each lattice, so most of the lattice volume is involved in computing
this correlator.
Also, the MILC code does a stochastic estimate of $\int d^4x\, \bar s s$
using a random source (covering the entire lattice) when the lattice is generated or 
read in for a measurement.  Thus,
we have $\bar s s$ measurements from several random sources on each lattice,
and can compute the correlation between the nucleon correlator and $\bar s s$.
(The number of random sources ranged from two to fifteen, depending on the
ensemble, with an average of ten.)
%In some of the recently extended ensembles, new nucleon correlators were
%computed on archived lattices.
% NOT NECESSARY EXCEPT FOR US

\begin{figure*}[tbh]
\hspace{-0.1in}
\includegraphics[width=0.325\textwidth]{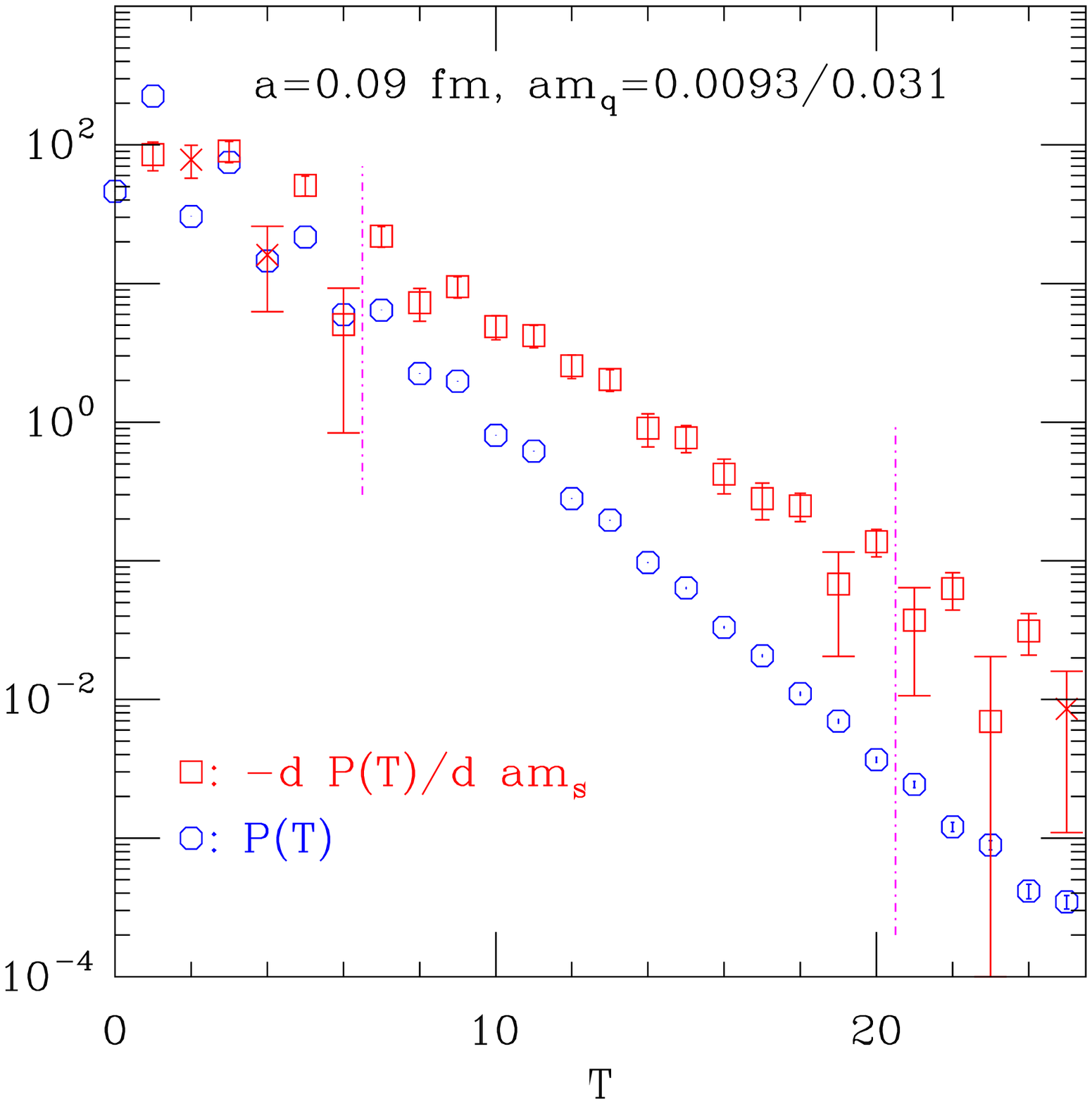}
\includegraphics[width=0.335\textwidth]{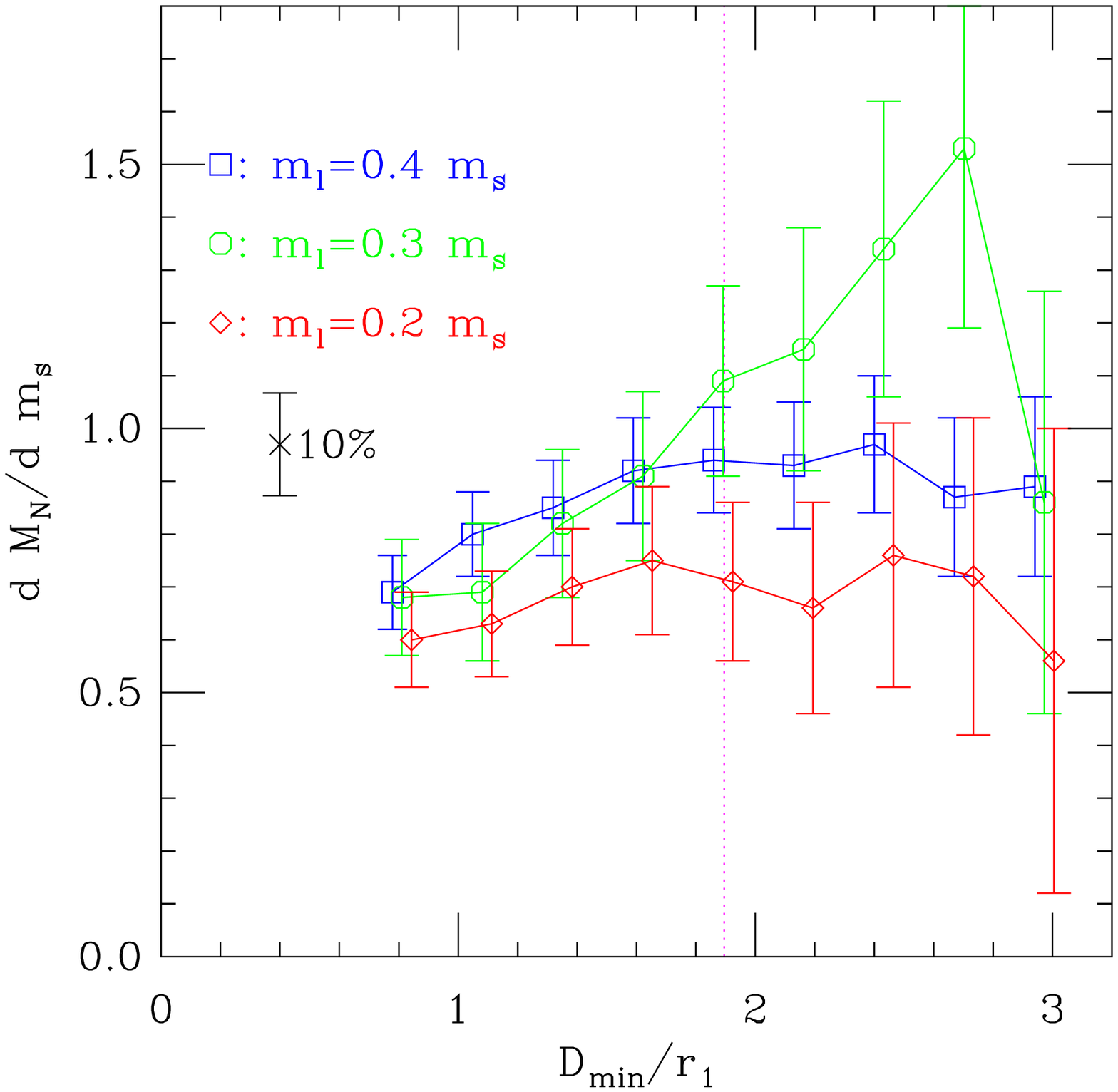}
\includegraphics[width=0.325\textwidth]{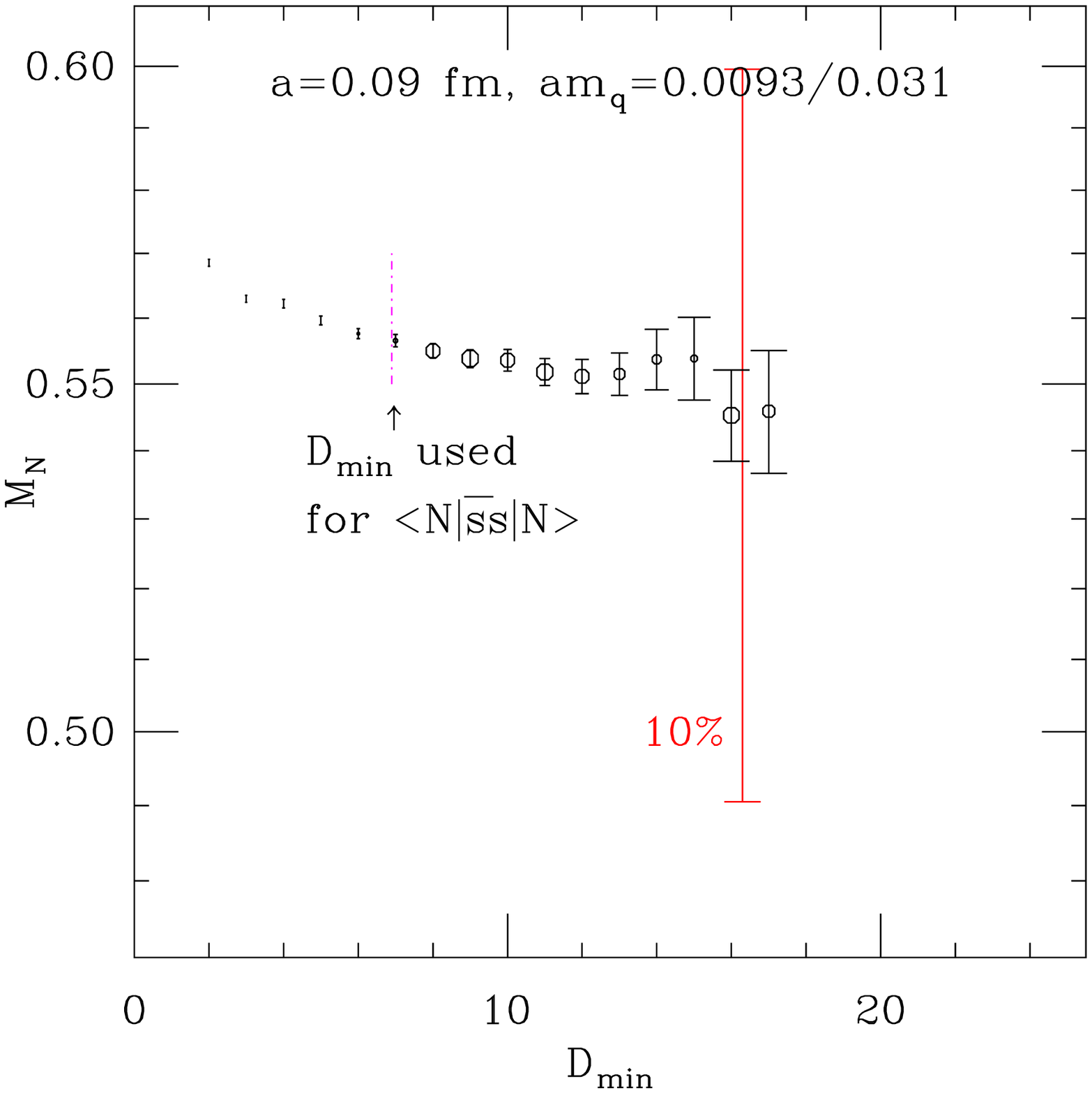}
\caption{The nucleon correlator and the derivative
of this correlator with respect to $m_s$ for the ensemble with
$am_l=0.0093$ and $am_s=0.031$ (first panel).  For the derivative, the squares
are points where the derivative is negative, and crosses are points
where it is positive.  The vertical lines show the range used in fitting the correlator.
The second panel shows $\PAR{M_N}{m_s}$ for three ensembles with $a\approx 0.9$fm as a
function of the minimum distance used in the fitting, and the third panel shows 
the fitted nucleon mass itself versus $D_{min}$.  The error bars labelled ``10\%''
in the second and third panels show the size of the ten percent
systematic error estimate from excited state contamination.
\label{samplefigs}
}
\end{figure*}

% choice of fit range, fit forms
We fit the nucleon correlators to a form including the nucleon and
an opposite parity state, using distance range $D_{min}$ to $D_{max}$
\BNE P(t) = A e^{-M_N t} + A^\prime (-1)^t e^{-M^\prime t} \ \ \ \ .\ENE
Since the fractional statistical errors on the nucleon correlator increase
quickly with minimum distance, it is advantageous to use as small
a minimum distance in the fits as possible.
Since we have a quark-line disconnected correlation function, statistical
errors are much larger than in simple hadron mass calculations.
Thus, in fitting the perturbed nucleon correlators,
we have chosen smaller minimum distances than in our fits to the nucleon masses
themselves.   In particular, we have chosen $D_{min}=5$, $7$ and $10$ for the $a=0.12$ fm,
$a=0.09$ fm and $a=0.06$ fm ensembles respectively, or a consistent physical
distance of about $0.6$ fm.  Since the nucleon mass, $M_N$, computed from these same
correlators can be determined with a
statistical error of order one percent, its dependence on minimum distance
can be used to estimate the resulting systematic error.
Fits to $M_N$ with these minimum
distances differ by between 1\% and 5\% from the $M_N$ fit
with larger minimum distances.  Alternatively, from looking at the values of $\PAR{M_N}{m_s}$
for various minimum distances, it appears that there could be errors as
large as 10\% from the choice of fit range.  (The choice of $D_{max}$
has negligible effects.)
% example graphs, nucleon propagator and its derivative
Figure~\ref{samplefigs} shows the nucleon correlator and its derivative
with respect to the strange quark mass for a sample ensemble.
The second panel of the figure shows $\PAR{M_N}{m_s}$ (unrenormalized)
for three of the $a \approx 0.09$ fm ensembles versus the
minimum distance included in the fit, while the third panel shows the
nucleon mass itself as a function of $D_{min}$.

% block 10 jackknife (larger blocks gave about same error)
Since the quantity we are computing is a complicated, and  implicitly defined,
function of the averages measured on the lattice, we use a jackknife
analysis to estimate statistical errors.  Since consecutive lattices
are correlated, we eliminated blocks of ten consecutive lattices, or 50
to 60 simulation time units, in the jackknife analysis.  Using larger
blocks made only a small difference.

%OLD \begin{figure}[tbh]
%OLD \includegraphics[width=0.3\textwidth]{propvalues_0093.ps}
%OLD \hspace{0.5in}
%OLD \includegraphics[width=0.3\textwidth]{sbsderiv_0093.ps}
%OLD \caption{The nucleon correlator (first panel) and the derivative
%OLD of this correlator with respect to $m_s$ (second panel) for the ensemble with
%OLD $am_l=0.0093$ and $am_s=0.031$.  For the derivative, the octagons
%OLD are points where the derivative is negative, and crosses are points
%OLD where it is positive.  The vertical lines show
%OLD the range used in fitting the correlator.
%OLD \label{samplefigs}
%OLD }
%OLD \end{figure}

% table of results

%\section{Renormalization}

% take from notes -> msbar
% explain why we do it first.
Since the strange quark mass is renormalization scheme and scale dependent, so is the
derivative $\PAR{M_N}{m_s}$.   For a useful result, we wish to express our
answer in a renormalization scheme useful for computations of cross sections.
%for interactions.  Fortunately, the relation between the strange quark
%dark matter interactions.  Fortunately, the relation between the strange quark
The relation between the strange quark
%mass in the Asqtad lattice regularization and the mass in the $\overline{\roman ms}$
mass in the Asqtad regularization and in the $\overline{\mathrm {MS}}$
scheme is known to two loop order in perturbation theory \cite{QUARKMASS2}.
%In particular,
\BNE \PAR{M_N}{m_s(\overline{\mathrm{MS}}, {\rm 2\ GeV})} = \frac{u_0}{Z_m} \PAR{M_N}{m_s({\rm Asqtad}, 1/a)} \ENE
where the factor of $u_0$ converts the lattice definition of the quark mass used
here to the definition used in Ref.~\cite{QUARKMASS2}, and $Z_m$ can be found in
Ref.~\cite{QUARKMASS2}.   Since in the subsequent steps in this analysis we will be
combining results at different lattice spacings $a$, it is most consistent to
make the conversion to the $\overline{\mathrm{MS}}$(2 GeV) scheme before making
chiral and continuum extrapolations.

% strange quark mass adjustment
%\section{Strange quark mass adjustment}

The strange quark masses, $m_s$, used in the MILC simulations were of necessity estimated
before the simulations were done, and the correct strange quark masses were
only known after the pseudoscalar masses were analyzed.  These differ significantly
from the $m_s$ used in the simulations.  For lattice spacings
$0.12$ fm, $0.09$ fm and $0.06$ fm the values of $am_s$ used in most of the simulations
were $am_s=0.050$, $0.031$ and $0.018$, while the corrected values are
$0.036$, $0.026$ and $0.019$, respectively.   (A few ensembles were run with a lighter
strange quark mass $0.6$ times the above mass.)   To adjust
to the correct $m_s$,
we use the fact that light quark $\bar\psi\psi$ was also evaluated on all of
these lattices.
%TRY (Of course, the correlation of this with the nucleon mass gives
%TRY the derivative of $M_N$ with respect to the light sea quark mass.)
In some of the largest ensembles the actual light quark mass used was still fairly
large --- 0.2, 0.4 or even 0.6 times the simulation strange quark mass,
outside the chiral regime and with qualitative behavior similar to
heavy quarks.
% REVISE add sentence below
For example, one of the ensembles with $a \approx 0.12$ fm was generated with
light and strange quark masses $am_l = 0.03$ and $am_s = 0.05$, where the correct
strange quark mass determined later was about $0.036$.
On these
ensembles we use the difference between $\PAR{M_N}{m_s}$ and $\PAR{M_N}{m_l}$
($m_l$ is the light quark mass used in the simulation) to calculate
the derivative of $\PAR{M_N}{m_s}$ with respect to $m_s$,
%$\PARTWO{M_N}{m_s}$
and use this to adjust the results to
the correct strange quark mass.   Since $\PAR{M_N}{m_s}$ and $\PAR{M_N}{m_l}$
are measured on the same lattices and with the same nucleon propagators, albeit
with different random sources, they are
highly correlated and the error on their difference is greatly
reduced. With the additional assumption that this slope in physical
units is the same for all ensembles, a correction factor can be estimated.
In particular, using five long ensembles with $m_l \ge 0.2 m_s$, we
find $\PAR{}{r_1 m_s} \LP \PAR{M_N}{m_s} \RP = -2.2(3)$.
Here $r_1$ is a hadronic length scale determined from the heavy quark
potential, and is approximately $0.31$ fm\cite{SOMMER,MILC_R1,RMP}.

% graphs of results, all ensembles, each lattice spacing
%\begin{widetext}
\begin{figure*}[tbh]
\hspace{-0.1in}
\includegraphics[width=0.302\textwidth]{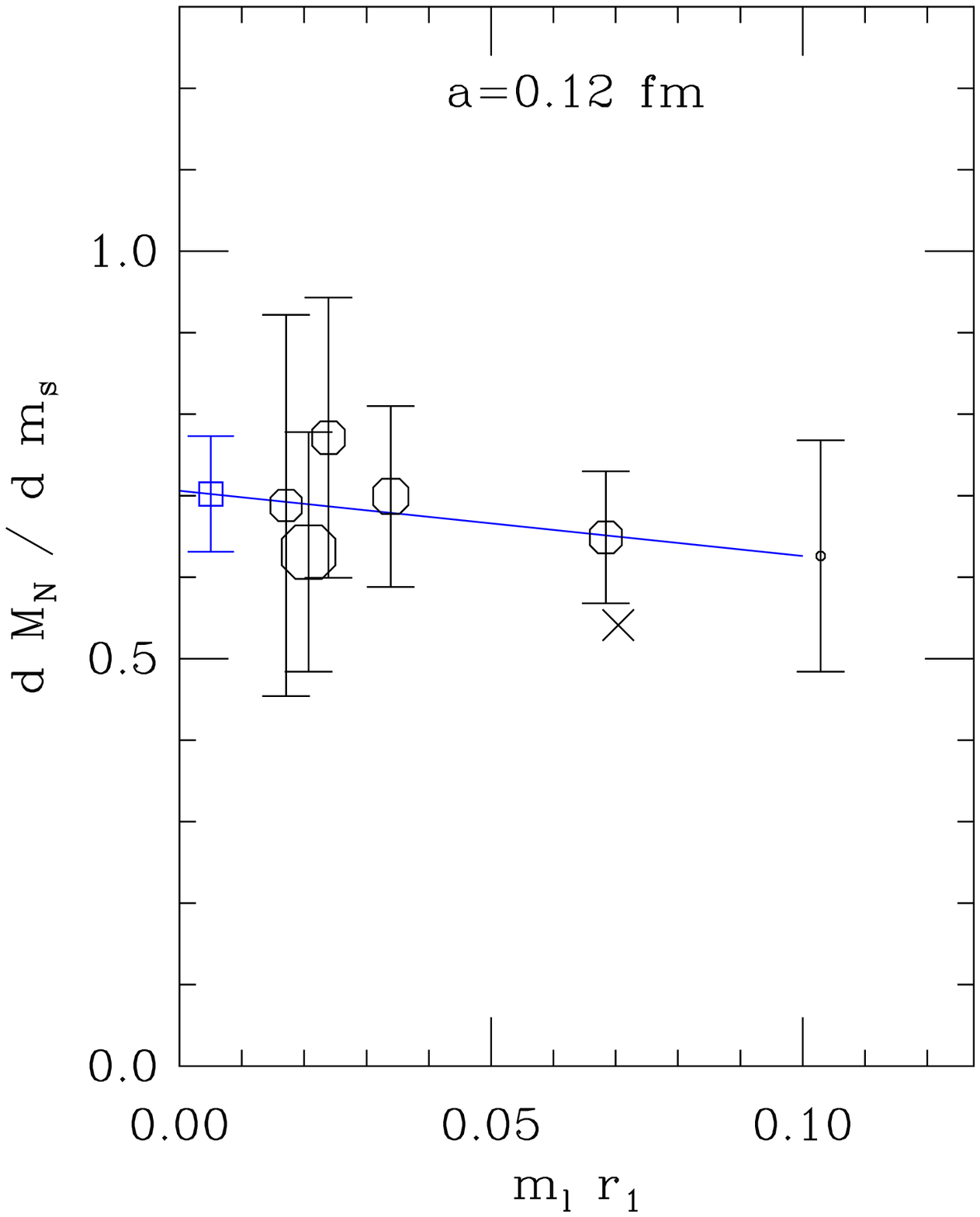}
\includegraphics[width=0.384\textwidth]{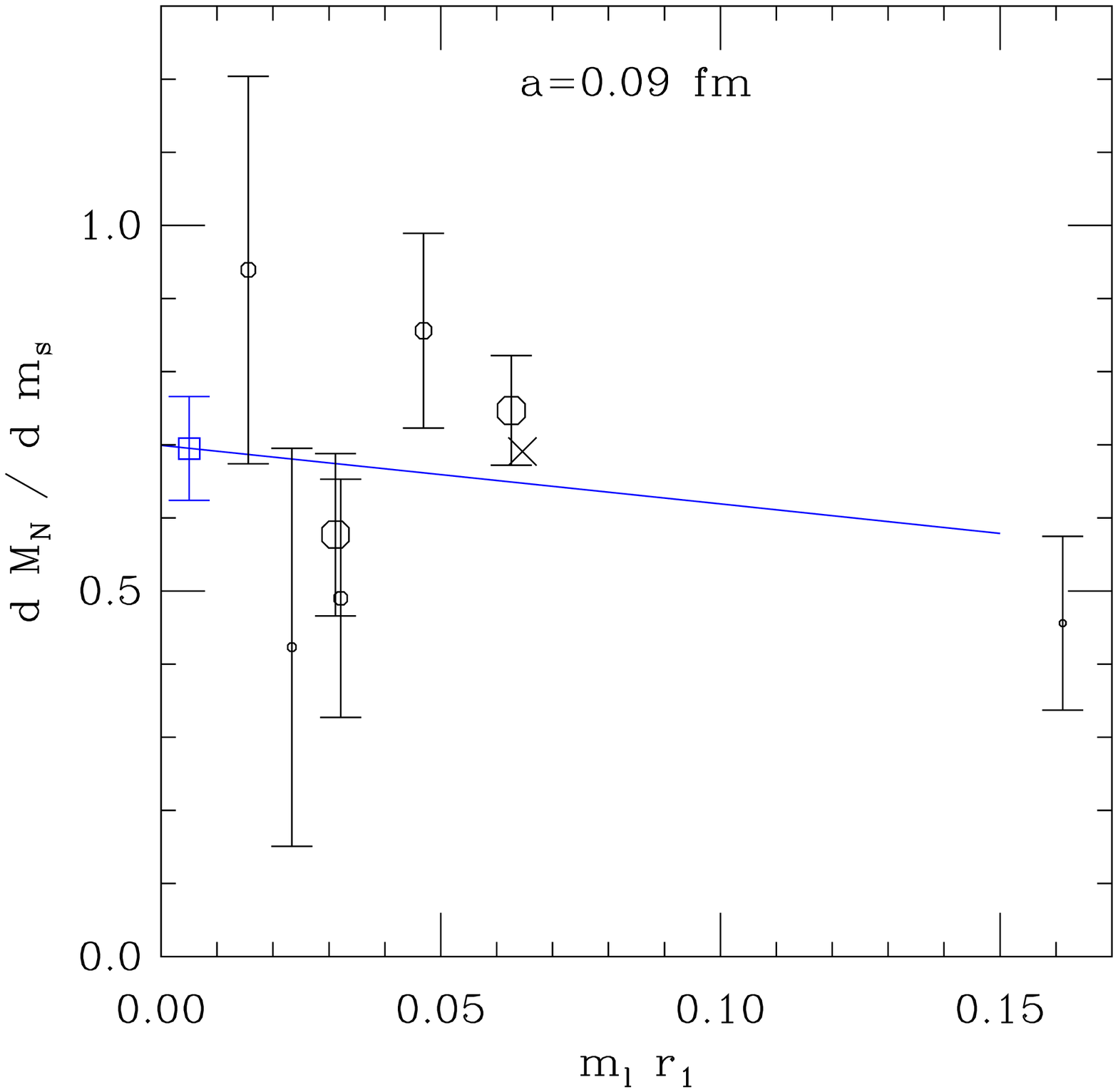}
\includegraphics[width=0.302\textwidth]{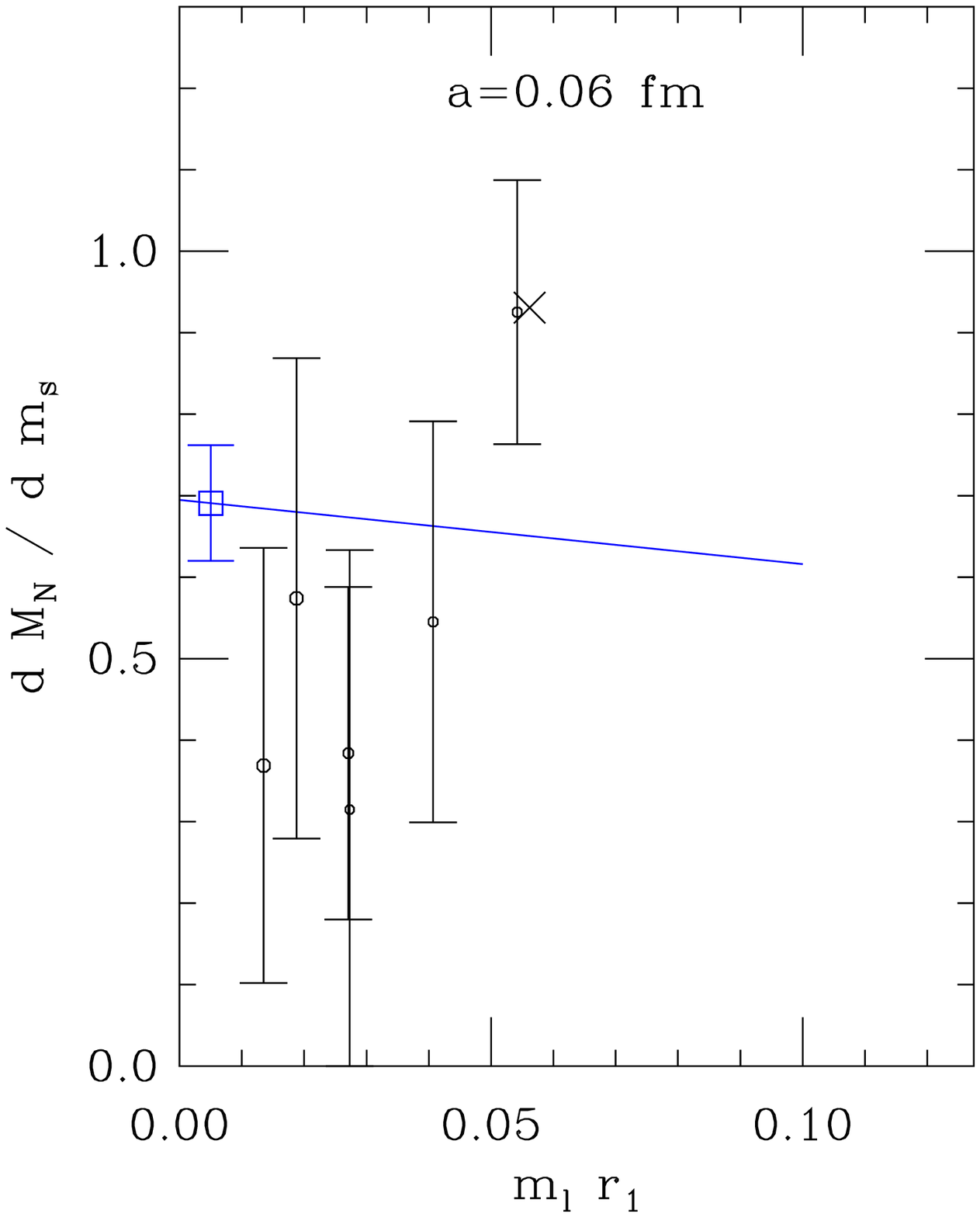}
\caption{The derivative $\PAR{M_N}{m_s}$ on the various ensembles.
As discussed above, the data have been adjusted to the correct strange
quark mass, and the quark mass converted to the $\overline{\mathrm{MS}}({\rm 2\ GeV})$
regularization.
%REVISE
%In the horizontal axis, $r_1$ is a hadronic length scale
%determined from the heavy quark potential, and is approximately $0.31$ fm.
In the horizontal axis, $r_1$ is a hadronic length scale,
approximately $0.31$ fm.
In these plots the symbol size is proportional to the number of lattices
in the ensemble, with the largest symbol corresponding to about 4500
lattices.
In each panel, the cross at $m_l r_1 \approx 0.05$ ($m_l \approx 0.4 m_s$) also shows
the value of the nearby point before adjusting the strange quark mass.
The line on each panel is the continuum and chiral fit in Eq.~\protect\ref{fitform}
evaluated at the corresponding lattice spacing, and the error bar at the left is
the error on the combined fit to all the data.
\label{resultfigs}
}
\end{figure*}
%\end{widetext}

Figure~\ref{resultfigs} shows $\PAR{M_N}{m_s}$ on all of the ensembles
used, where the results have been adjusted to the correct strange
quark mass, and the quark mass converted to the $\overline{\mathrm{MS}}({\rm 2\ GeV})$
regularization.
In this plot we see that the best results are in the $a=0.12$ fm ensembles,
mainly because of the larger numbers of lattices.

%\section{light quark chiral limit}
%It is clear in Fig.~\ref{resultfigs} that only the $a=0.12$ fm data strongly
%constrains the dependence on light quark mass.  We therefore use the slope
%of the linear fit in the first panel of this figure as our estimate of
%the linear term in the chiral extrapolation, and the increase in variance
%of the result at the physical quark mass when the fit is changed from constant
%to linear as an estimate of the error on $\PAR{M_N}{m_s}$ from the uncertainty
%in this slope.  To estimate the effects of higher order terms in chiral perturbation
%theory, 
%\\FIXX THIS

%\section{Continuum limit}
% fit using Bayesian priors
Finally, it is necessary to extrapolate the result to the physical light
quark mass and to the continuum ($a=0$) limit.
To do this, we fit the results to the form\cite{FRINKMEISSNER}
\BNE\label{fitform} \PAR{M_N}{m_s} = A + B m_l r_1 + C (a/r_1)^2\ \ \ .\ENE
Since the results from the $a=0.06$ and $0.09$ fm ensembles have much larger
statistical errors than the $0.12$ fm results, the term linear in $a^2$ is
very poorly determined.
However, we can use experience with other quantities
to estimate the likely size of lattice corrections.  In particular, the masses of
the $\rho$, nucleon, and $\Omega^-$
%(in units of $r_1$)
at $a=0.12$ fm differ by about 4\%,
10\% and 9\% respectively from their continuum extrapolation.
Therefore we constrain $C$ to be small by using a (Gaussian) Bayesian prior
with a one standard deviation width corresponding to a 10\% effect at $a=0.12$ fm.
This gives $\PAR{M_N}{m_s} = 0.69 \pm 0.07_{\rm statistical}$ in the continuum limit,
with $\chi^2/D = 17.0/17$.

There are also a number of systematic errors.
%TRY Table~\ref{errortable} summarizes the errors we have considered.
As discussed above, we include a 10\% systematic
error for the effects of excited states in the nucleon correlator.  The 
extrapolation to the physical light quark mass contains higher order terms
in chiral perturbation theory
than the linear form used here.   To estimate the likely size of these
terms, we note that if the nucleon mass over this range of quark masses
is fit to constant plus linear, the result at the physical point is
seven percent different from the result including two more orders in the
pion mass.  We therefore take seven percent as an estimate
of the effect of higher order terms in chiral perturbation theory.
In one case where we have two spatial volumes, the nucleon mass on the
volume used here was different by about one percent from the mass in the
larger volume.   It is possible that disconnected contributions are more
sensitive to the volume, so we take three percent as an estimate of this
systematic error.
%Finally, the authors of Ref.~\cite{QUARKMASS2} estimate
%a remaining error of four percent in $Z_m$.
Finally, Ref.~\cite{QUARKMASS2} estimates an error of four percent in $Z_m$.
The combined systematic error estimate from excited states, finite volume,
higher order $\chi PT$ and $Z_m$ is 0.09.

Evaluating the fit in the continuum limit at the physical light
quark mass, we find $\PAR{M_N}{m_s} = 0.69 \pm 0.07_{\rm statistical} \pm 0.09_{\rm systematic}$, where
$m_s$ is in the $\overline{\mathrm{MS}}$ regularization at 2 GeV.
It is also common to quote the renormalization scheme invariant quantity
$m_s \PAR{M_N}{m_s}$.  Using a similar chiral and continuum fit to the one used
for $\PAR{M_N}{m_s}$, we find $m_s \PAR{M_N}{m_s} = 59(6)(8)$ MeV.
%The systematic error here does not include error in $Z_m$, since this cancels, but does
The systematic error here does not include error in $Z_m$, which cancels, but does
include a lattice systematics error of almost the same amount, coming from
uncertainty in the lattice strange quark mass and an overall two percent
error in scale setting.

%\section{Systematic errors}
% finite volume is 3% here, note estimated nucleon mass off by 1%
%TRY \begin{table}[tbh]
%TRY \begin{tabular}{ll}
%TRY \hline
%TRY Source & Estimate \\
%TRY \hline
%TRY statistical &  0.070 \\
%TRY \hline
%TRY Excited states & 0.069 (10\%) \\
%TRY Finite volume & 0.021 (3\%) \\
%TRY Higher order $\chi PT$ & 0.049 (7\%) \\
%TRY Error in $Z_m$  & 0.028 (4\%) \\
%TRY Combined systematic & 0.09 \\
%TRY \hline
%TRY \end{tabular}
%TRY \caption{
%TRY \label{errortable}
%TRY Error budget.   The justification for each estimate is in the text.}
%TRY \end{table}
% revise?  We could drop the table and just add sentence saying
% the above systematic errors combine to 0.09

% note on partially quenched check
% 5/4/09 Walter's updated version
In general the MILC ensembles were run at different lattice coupling for each
quark mass, which makes it complicated to extract $\PAR{M_N}{m_q}$ from fits
to the table of nucleon masses.  However, in one case there is an accidental
check.  Through an error, two ensembles were run with the same coupling constant
$10/g^2$ and tadpole factor $u_0$.
These ensembles had sea quark masses $m_l/m_s = 0.0062/0.0186$ and
$0.0093/0.031$ respectively. By computing a partially quenched nucleon mass 
on the latter ensemble and examining its difference from the nucleon mass on 
the former, we can make a check on a particular combination
of $\PAR{M_N}{m_s}$ and $\PAR{M_N}{m_l}$,
$(0.031-0.0186)\PAR{M_N}{m_s}+2(0.0093-0.0062)\PAR{M_N}{m_l} $.
%= m_{N,m_l=0.0093,m_s=0.031} - m_{N,m_l=0.0062,m_s=0.0186}$.
Here $\PAR{M_N}{m_q}$ is evaluated at the midpoint of the sea quark masses
on these two ensembles, and the factor of two comes from the
two light flavors.
These nucleon masses are computed in the usual way by a fit to the nucleon correlator.
%The resulting difference in masses was $0.016 \pm 0.003_{\rm stat.} \pm 0.002_{\rm fit\ range} = 0.016 \pm 0.004$.
The resulting difference in masses was $0.016(3)_{\rm stat.}(2)_{\rm fit\ range} = 0.016(4)$.
The fit to $\PAR{M_N}{m_s}$ above, converted
back into lattice units, together with a similar fit to $\PAR{M_N}{m_l}$,
%gives $0.020 \pm 0.003$, in reasonable agreement with this check.
gives $0.020(3)$, in reasonable agreement.
% see klingon:~doug/Asq/3flav_fitting/ssbarnuc/ANALYSIS for details
% If we end up deleting this section, we need to delete the reference to it in
% the introduction.

%TRY The light quark content of the nucleon, $\LL N | (\bar u u + \bar d d ) | N \RR$, is also
%TRY interesting.  The sea quark contribution to this quantity can be determined by similar
%TRY methods to those used here.  Because the statistical errors from the light quark
%TRY $\bar\psi\psi$ are larger, and because the chiral extrapolation is more difficult,
%TRY we defer discussion of this to later work.
%OLD
% Our result  for $\LL N | \bar s s | N \RR$ is smaller than the results of the quenched calculations in 
% Refs.~\cite{FUKUGITA02,DONG02}.
% However, with commonly quoted
% values for $\sigma_N$ our result is consistent with
% the small value of $y$ recently found in the two flavor overlap
% calculation in Ref.~\cite{JLQCD08}.
% On the other hand, our result is larger than the result from fits to
% baryon masses in Ref.~\cite{YOUNGTHOMAS09}, likely because of differences
% in how the derivative with respect to $m_s$ is taken.
% %whose result can be
% %expressed as $\PAR{M_N}{m_s}=0.22 \pm 0.12_{\rm statistical} \pm 0.08_{\rm systematic}$.
%NEW
Our result  for $\LL N | \bar s s | N \RR$ is smaller than the results of the quenched calculations in 
Refs.~\cite{FUKUGITA02,DONG02}.
However, our result is reasonably consistent with
the small value of $y$ recently found in the two flavor overlap
calculation in Ref.~\cite{JLQCD08}, where combining their result $y<0.05$ with the
value of $\sigma_{\pi N} = 53$ MeV found in the same fit gives $m_s<\bar s s> < 36$ MeV.
Similarly, our result is marginally consistent with the result from fits to
baryon masses in Ref.~\cite{YOUNGTHOMAS09}, who find $m_s<N|\bar s s|N> = 31(15)$ MeV,
although there may be differences in how the derivative with respect to $m_s$ is taken.
%whose result can be
%expressed as $\PAR{M_N}{m_s}=0.22 \pm 0.12_{\rm statistical} \pm 0.08_{\rm systematic}$.

\vspace{-0.2in}
\section*{Acknowledgements}
\vspace{-0.25in}
This work was supported by the U.S. Department of Energy grant number DE-FG02-04ER-41298.
Additional computation for this work was done at the Texas Advanced Computing
Center (TACC) and the National Energy Resources Supercomputing Center (NERSC).
We thank the UKQCD collaboration for providing some of their lattices.
We thank Alexei Bazavov, Claude Bernard, Carleton Detar, Craig McNeile,
James Osborn and Bira van Kolck for helpful suggestions and assistance with the
UKQCD lattices.

\end{document}